# First Tests of Gaseous Detectors Made of a Resistive Mesh

R. Oliveira[1], V. Peskov[1,2], F. Pietropaolo[3], P. Picchi[4]
[1]CERN, Geneva, Switzerland,
[2]UNAM, Mexico City, Mexico
[3]INFN Padova, Padove, Italy
[4]INFN Frascati, Frascati, Italy

**Abstract**

We describe here various detectors designs: GEM-type, MICROMEGAS-type, PPAC-type as well as cascaded detectors made of a resistive mesh manufactured from a resistive Kapton foil, (20μm thick, resistivity a few MΩ/□) by a laser drilling technique. As in any other micropattern detectors the maximum achievable gas gain of these detectors is restricted by the Raether limit, however, the resistive mesh makes them and the front end electronics fully spark protected. This approach could be an alternative/or complimentary to the ongoing efforts in developing MICROMEGAS and GEMs with resistive anode readout plates and can be especially beneficial in the case of micropattern detectors combined with a micropixel-type integrated front end electronics.

**1. Introduction**

Due to their very high granularity, recently developed micropattern gaseous detectors offer new possibilities in experimental techniques and applications, and for this reason are the subjects of a great interest. However, the fine structures of their electrodes make the micropattern detectors very fragile, for example they can be easily damaged by occasional sparks. Studies show (see [1] and references therein) that sparks appear when the total charge in avalanche exceeds some critical value $Q_{crit}$:

$$A_m n_0 = Q_{crit} \quad (1),$$

where $A_m$ is the maximum achievable gain before the breakdown appear and $n_0$ is the number of primary electrons produced in the detector volume by the ionizing radiation. The physic behind this is that at this critical value of $Q_{crit}$ the space charge in the avalanche becomes sufficient to disturb the external electric filed. As a result, photoelectrons created by avalanches in the surrounding gas volume begin to drift towards the positive ions remaining from the initial avalanche and finally form a thin plasma filament, called a streamer. Typically, for micropattern detectors depending on their design and operational pressure

$$Q_{crit} = 10^6\text{-}10^7 \text{ electrons} \quad (2).$$

It should be stressed that in practical situations to avoid an excessive number of potentially damaging sparks, one has to work at gas gains of one or two orders of magnitude smaller than the gains that correspond to the $A_m$ estimated from the formula (1).Thus, if micropattern detectors are used to only detect radiation producing relatively small amount of primary electrons $n_0 \sim 100$ (minimum ionizing particles, soft X-rays) they can reliably operate at gains $\leq 10^3$. However, in the presence of heavily ionizing particles and/or neutron background producing dense ionization with a very high value of $n_0$, the maximum achievable gain will correspondingly drop.

Actually, the main problem appears during the long term operation of micropttern detectors: even if they operate at relatively low gains at some moment a particle producing high values of $n_0$ may appear (for example, due to the natural radioactivity) and this may trigger the breakdown. Moreover, there is also some probability of avalanches overlapping in time and the space as well sporadic emissions of electron jets from the cathodes - another source of breakdowns(see [1].As a result during the long-term operation of micropattern detectors, even at gas gains of $A \ll A_m$, occasional sparks are almost unavoidable. Coming from this theoretical and experimental facts the "detector community" now accepts that in real experimental conditions (for example, during a long term high energy physics experiment) GEMs and MICROMEGAS will always have some probability of sparking and the main efforts are focused on reducing this probability as much as possible and on measures of protecting detectors from damage. These measures include: segmentation of electrodes on several isolating parts (to reduce the capacitance involved in the discharge process), use of cascaded detectors ( this increases $Q_{crit}$ [1]) and of course when it is possible, to use of spark-protected front end electronics. Unfortunately, these efforts had a limited success so far. For example, in cascaded detectors there could be discharge propagations from on element to another one and to the readout plate; this effect has higher probability in the presence of heavily ionizing particles [1]. Note that in case of the detectors combined with a multipixel SMOS array or MediPix (see [2, 3]) the requirements for the spark protection are exceptionally high due to the fragility of the readout system.

Recently we suggested another more radical solution for the spark protection of micropattern detectors: the use of resistive electrodes instead of traditional metallic ones (see for example [4] and references there in). This approach was successfully implemented in the case of GEM-like detectors having a thickness of >0.2mm (thinner GEMs were not possible to produce by the mechanical drilling technique we used).We named this detector a Resistive Electrode Thick GEM or RETGEM. Resistive electrodes reduce the energy released by the sparks by 100-1000 times and thus make these detectors as well as the front end electronics fully spark-protected. Nowadays, several groups experimenting with such detectors and they fully confirmed our earlier results [5-7]. Unfortunately, the position resolution of RETGEM is restricted by the hole 's diameter and the pitch and is about 0.7mm. Coming from this promising experience obtained with RETGEM, some other groups begun investigating the possibility to applying the resistive electrode approach to MICROMEGAS; prototypes of such detectors with anodes coated with various resistive layers were already developed and tested [8]. Although these studies are still in progress, the preliminary results clearly indicate that spark protection can be achieved by this technique.

In this report we will present an alternative /complimentary approach: detectors made of thin resistive meshes (Resistive Mesh Detectors or RMDs). From such resistive meshes one can assemble various detectors: GEMs with gaps below 0.2mm, MICROMEGAS, PPAC/RPC (including large gap PPAC/RPC ) as well as multistep detectors operating in cascaded mode.

## 2. Materials and Methods.

2.1 Resistive Meshes and Detector Design

The resistive meshes were manufactured from the resistive Kapton (resistivity 2.8-3 M$\Omega$/□) used earlier by us in the RETGEM designs [9]. The resistive meshes had the following geometries: mesh #1 had a thickness of t= 20μm, hole's diameter d=70 μm and hole spacing a=140 μm, it was made by a laser drilling technique[*](see Fig. 1); mesh #2 had t=25 μm, d=0.7 mm and a=1.7 mm; mesh #3 had t=25 μm , d=0.8 mm, a=2.8mm; meshed # 2 and #3 were manufactured by usual mechanical drilling techniques.

The resistive meshes were stretched either on 5x5 cm$^2$ or 10x10 cm$^2$ G-10 frames. From these stretched meshes different detectors could be assembled: PPAC/RPC (RM-RPC) with an avalanche gap G=1-3 mm (see Fig. 2a), resistive mesh MICROMEGAS (RM-μM), G=0.1-0.3 mm (Fig. 2b), resistive mesh GEM (RM-GEM), G=0.05-0.1mm (Fig. 2c) or several cascaded RM-GEMs or RM-GEM combined with RM- μM (as shown in (Fig. 2d). As anodes we used ether metallic plates or G-10 plates covered with resistive Kapton, or (only for position resolution measurements) a ceramic plate with a 50 pitch metallic strips. As a spacer between the meshes and the anode plate or between two meshes we used a plastic honeycomb structure (for PPAC/RPC) or 50-300 μm thick fishing lines, or 50 μm thick meshes made of usual Kapton (for RM-GEMs and RM- μM).

As an example in Fig 3a and b photographs of PPAC/RPC are presented made from the resistive mesh #3.

In some tests one of the resistive meshes was coated with an 0.4μm thick CsI layer.

2.2. Experimental Setup

The experimental setup consisted from a gas chamber housing one of the mentioned above RMDs, an Ar (Hg) UV lamp (in position resolution measurements a pulse D$_2$ lamp also was used), a monochromator, a lens focusing the light from the lamp to the input slit of the monochromator (or directly to the top RETHGEM surface (for position-resolution measurements) and a gas system allowing one to flush various gases: Ne- or Ar- based mixtures with various percentages of CH$_4$ or CO$_2$, or pure CH$_4$ (see Fig. 4).The ionization inside the gas

---
[*] A similar mesh was independently developed and is under the study of I. Laktineen (Lyon)[10]

volume has been produced by 5.9 keV photons from a $^{55}$Fe source or by alpha particles from an $^{241}$Am source. The signals from the detectors were detected by standard laboratory electronics- a charge sensitive amplifier Ortec 142PC. In some cases an amplifier Ortec 572 was also additionally used. With the help of the scope Le Croy LC 564A we could obtain a pulse height spectrum of the pulses or if necessary convert them to NIM standard pulses and count with a scaler CAEN N145. In the case of the position resolution measurements (when the ceramic anode plate was used) strips in the central region of the plate were interconnected together in groups (two in each) and six such groups were connected each to its own Ortec 142pc preamplifier.

In measurements with sparks a current amplifier was used (see [11] for more details).

For the calibration of the absolute intensity of the light beam exiting the monochromator a single- wire counter was used flushed with a mixture of $CH_4$ with TMAE vapors at temperature of 30°C(see Fig. 5). The active region of this detector was about 4 cm ensuring full UV light absorption inside its volume.

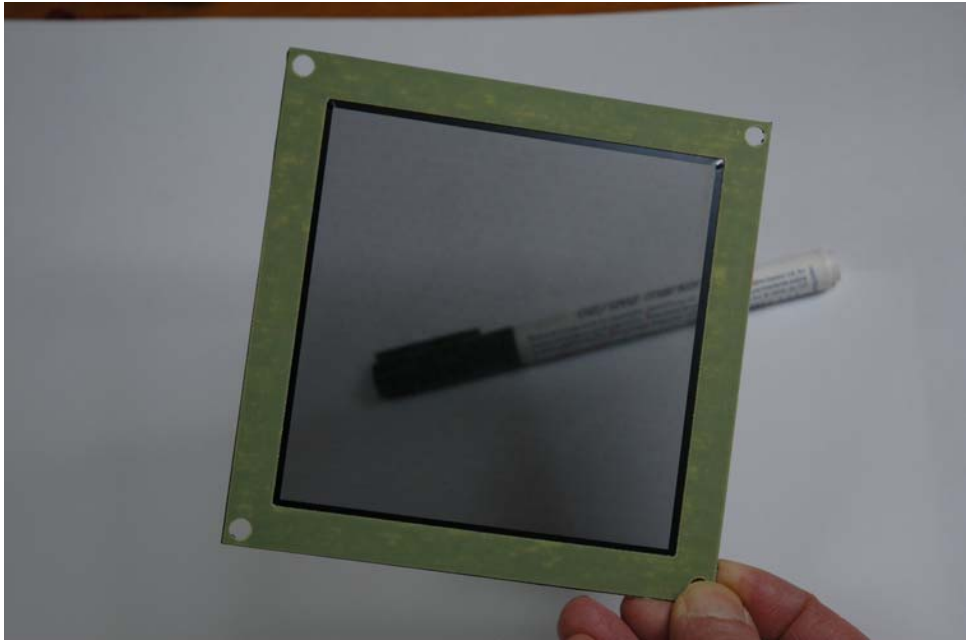

Fig. 1. A photograph of a resistive mesh manufactured by the laser drilling technique

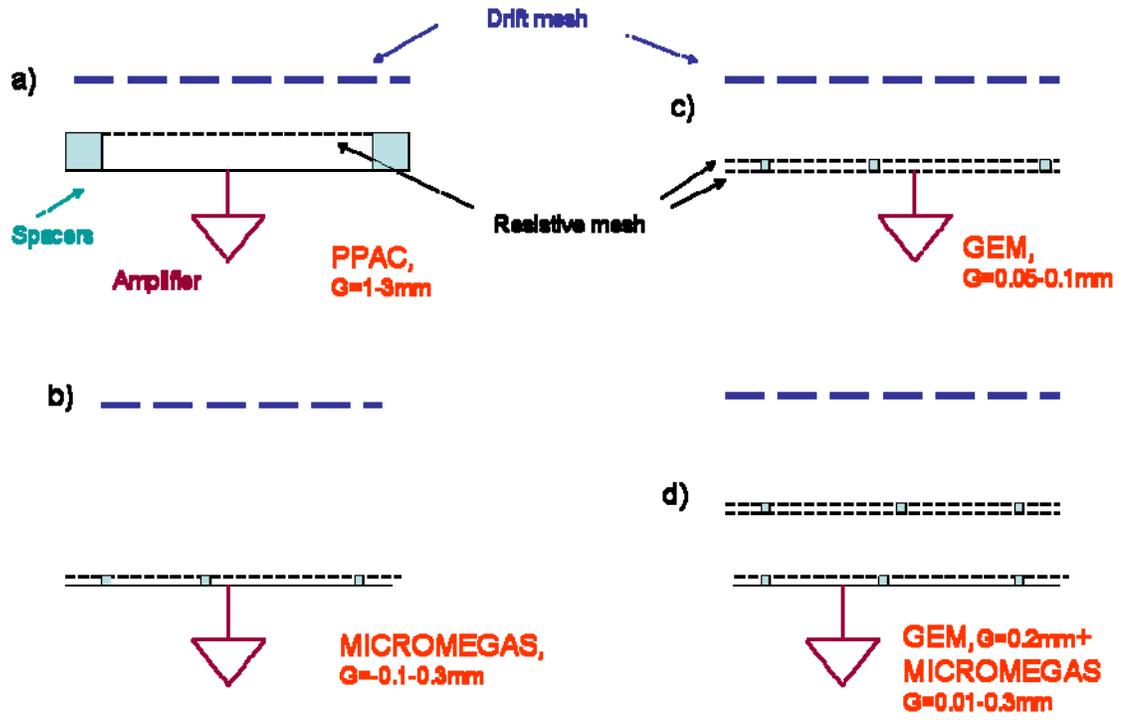

Fig. 2. Various designs of resistive mesh-based detector tested in this work

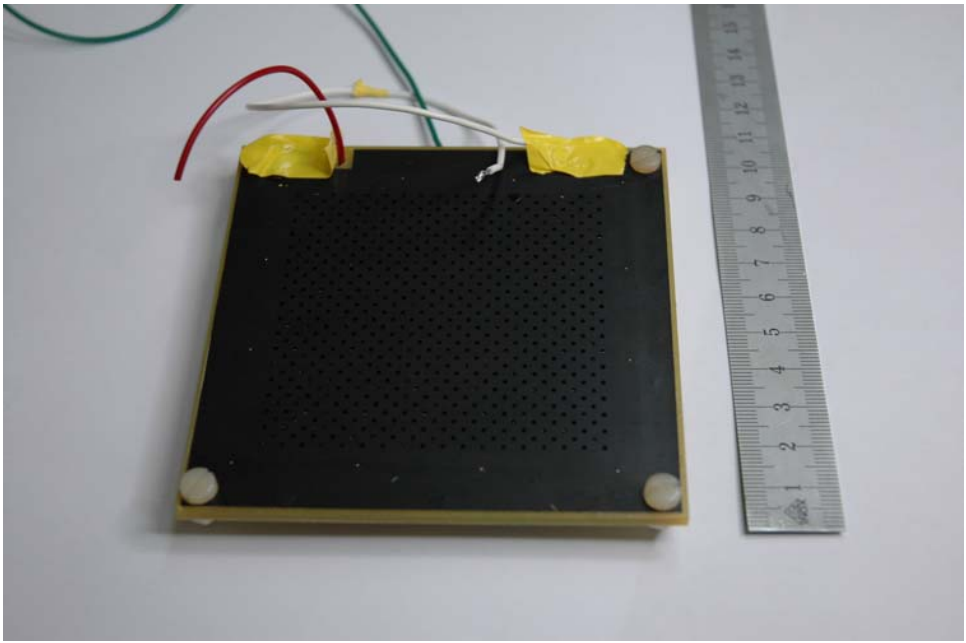

a)

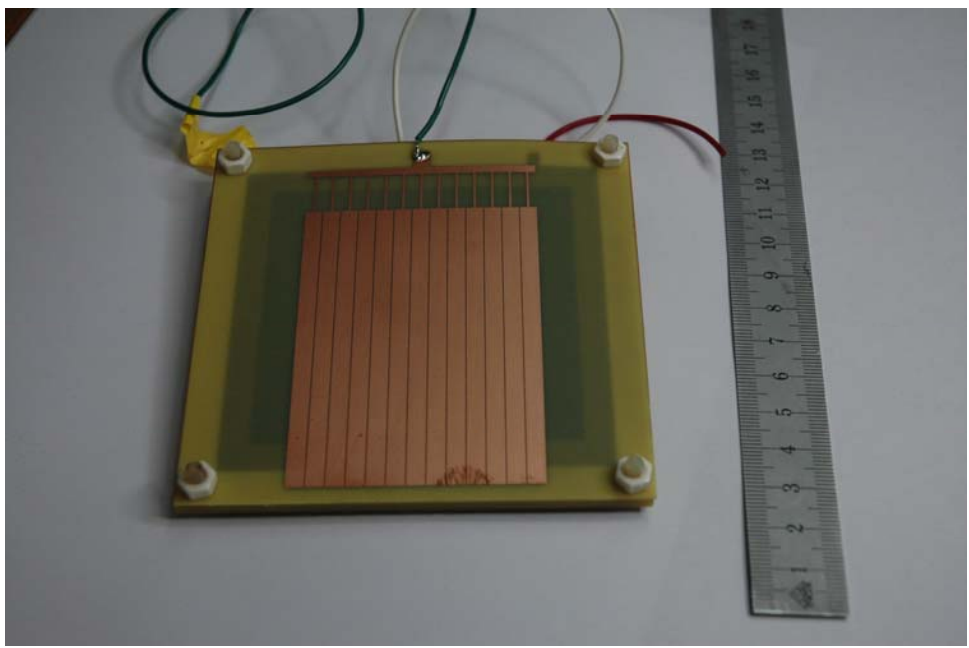

b)

Fig.3. Photographs of the PPAC/RPC (G=3mm) constructed from the mesh #3: a) the top view where the resistive mesh is clearly seen, b) the bottom view showing the readout strips

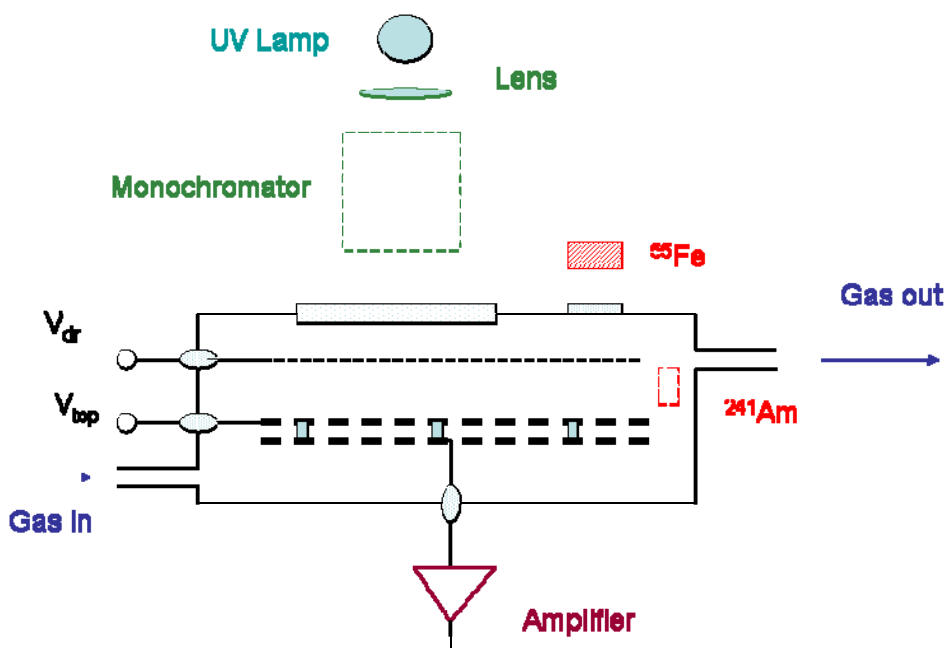

Fig. 4. A schematic drawing of the experimental setup

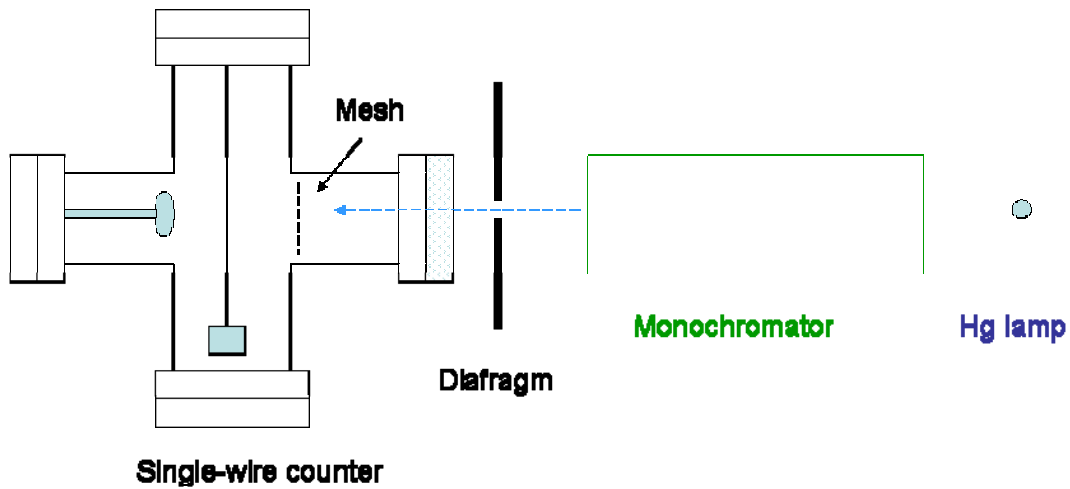

Fig.5. A schematic drawing of a single- wire counter used for the calibration of the absolute intensity of the UV beam from the monohromator

**3. Results**

3.1 Bare Resistive Meshes

Figs. 5 and 6 show the gain vs. voltage curves measured for RM-RPCs in Ne- and Ar-based mixtures. The measurements were stopped at voltages when first breakdown appear. As can be seen, with alpha particles the maximum achievable gains were around 100 which roughly corresponds the value expected from the Raether limit –see (2).
With $^{55}$Fe gas gains ~$10^4$ were reached with designs having no inner spacers between the resistive mesh and the metallic anode plate. This is ~10 times below the value expected from the Raeter limit and we attribute the origin of these breakdowns to imperfection of the mesh surfaces around the holes. Note that in the presence of the spacers the maximum achievable gain significantly dropped. In this case at voltages close to the breakdown values we could clearly observe from the scope the appearance of pulses having very high amplitudes similar as it was recorded earlier with the microstrip detectors [12]. We thus concluded that the streamers across the spacers surfaces are responsible for this maximum achievable gain drop and thus optimization of the spacers shape is essential in achieving high gains. The energy resolution obtained in Ar+$CH_4$ with $^{55}$Fe radioactive source was about 30-35 %. In Fig. 7 is depicted the signal amplitude vs. the counting rate for RM-RPC, G=1mm with metallic anode. The measurements were performed up to counting rate of 4.7 kH/cm$^2$ (the maximum counting rate available from our $^{55}$Fe source); no gain drop was observed so far.

The conclusion that in measurements with $^{55}$Fe the maximum achievable gain was limited by the charge leak across the spacers was fully confirmed during experiments with small-gap RMDs. Gas gains obtained with RM-µM (G=0.1-0.3mm) were rather low (below 10) in the case of Kapton spacers, however in the case of the fishing line spacers, the maximum achievable gains were close to $10^2$ (see Figs. 8 and 10.)

The gas gain achieved with RM-GEMs (G=0.05-0.01mm) made from two parallel resistive meshes separated by Kapton spacers were not high (see Fig.11), most probably due to the Kapton spacers and mesh defects. However, RM-µM combined with RM-GEM based preamplifier exhibited an overall gain close to $10^4$ (Fig.12).

The discharge current measurements performed as described in work [11], revealed that the resistive mesh reduces the spark current at least 10 times and reduces the energy released by the spark by 100 times (exact value depended on a particular design) compared to the similar results obtained with metallic mesh detectors.

It is important to note that even at very high transfer electric filed between the RM-GEM an RM- µM (> 2 kV/cm) there was no discharge propagation between the detectors. The obvious explanation is that this is due to the strong suppression of the discharge energy in the resistive mesh-based detectors and consequently, the reduction of spark's UV emissions which are responsible for the discharge propagation [1]

The main conclusion from these studies is that the resistive meshes are convenient constriction blocks for various designs of spark protected detectors.

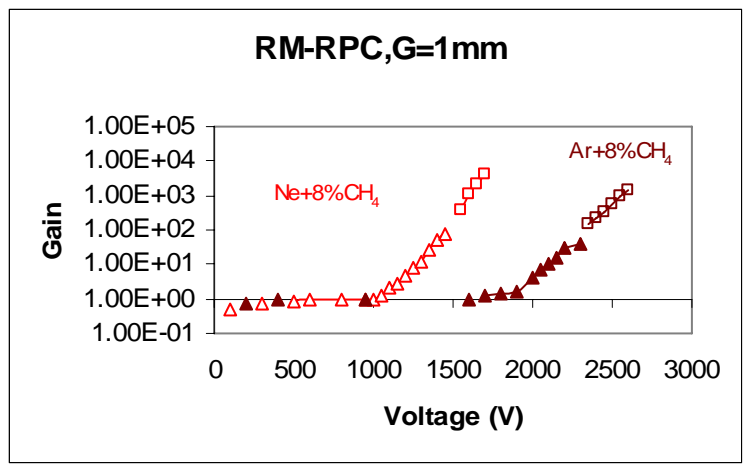

Fig. 5. Gain curves for 1mm RM-RPC (metallic anode, no spacers) measured in Ne and Ar mixtures with 8% of CH$_4$. Triangles- results of measurements performed with alpha particles, squares –measurements with $^{55}$Fe

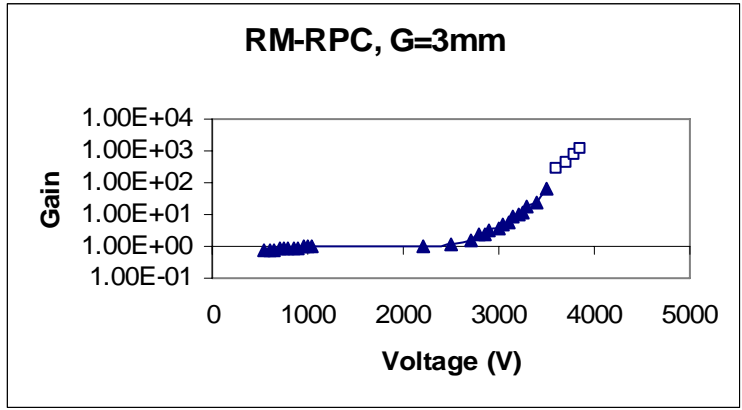

Fig 6. The gain vs. voltage measured for the RM-RPC(resistive Kapton anode, no spacers, G=3mm) in Ne+ 12%$CH_4$. Triangles-represent the results of measurements with alpha particles, squares -$^{55}$Fe

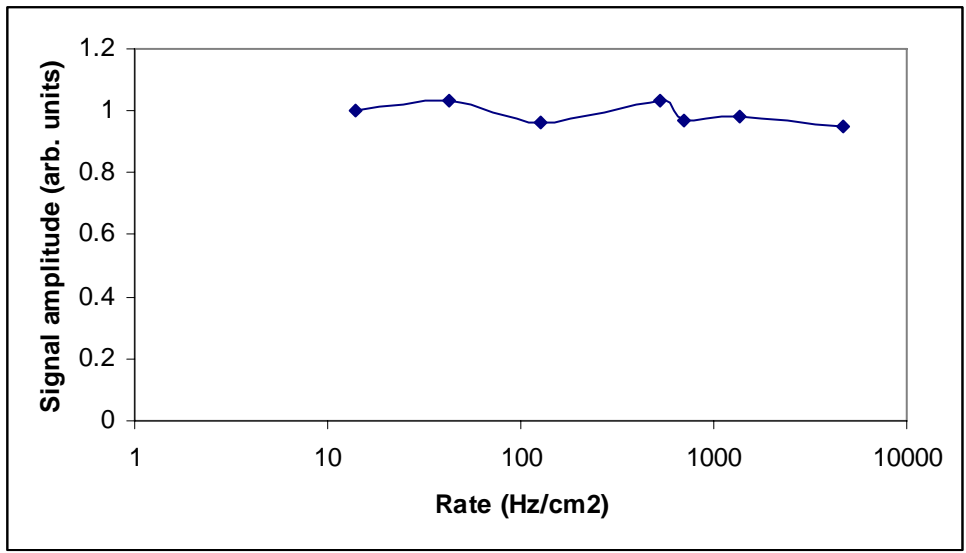

Fig.7. Rate characteristics of RM-RPC (G=1mm, metallic anode) measured with $^{55}$Fe source at a gas gain of 2000 in Ne+8%$CH_4$

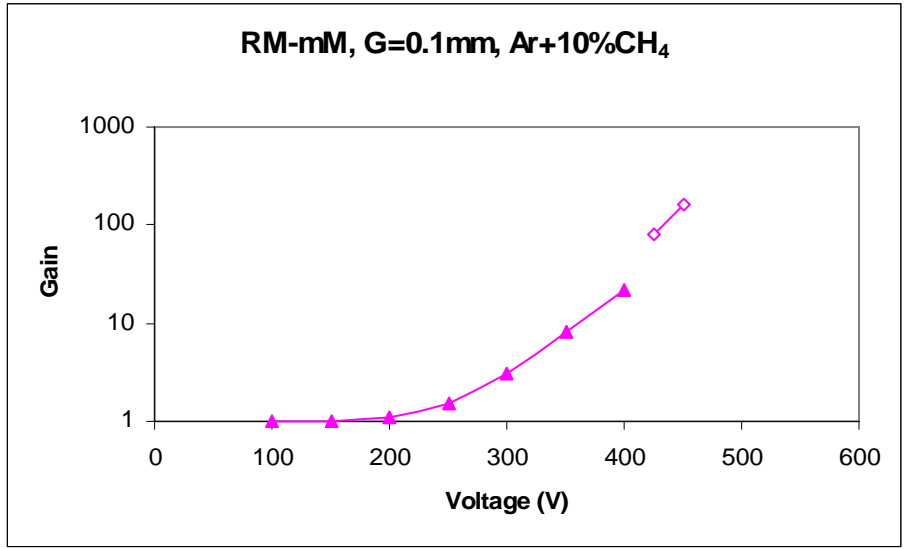

Fig.8. The gas gain variation with the anode voltage for RM-μM, G=0.1mm (fishing line spacers) operating in Ar+10%CH$_4$. Triangles-alpha particles, rhombuses- 60 keV X-rays from $^{241}$Am

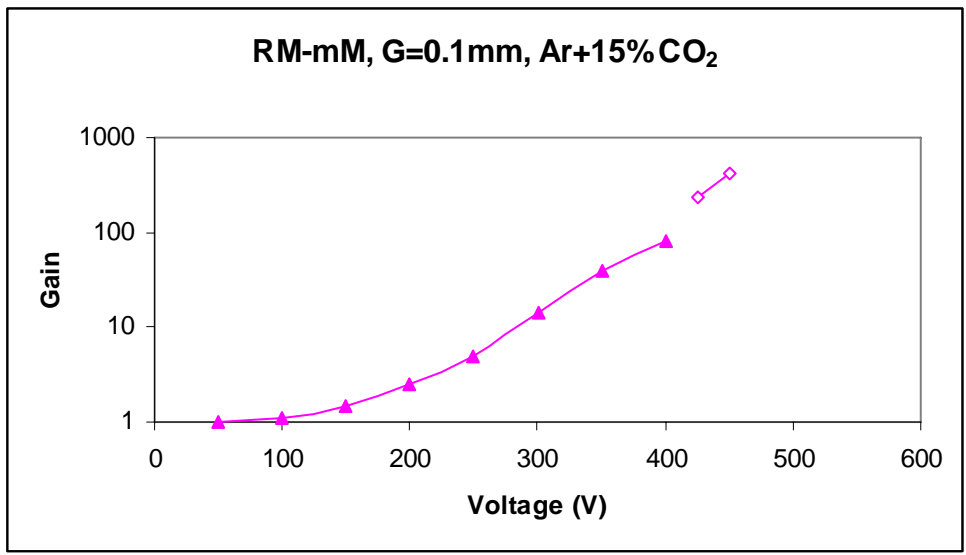

Fig.9. Gain measurements with the same RM-μM, but performed in Ar+15%CO$_2$

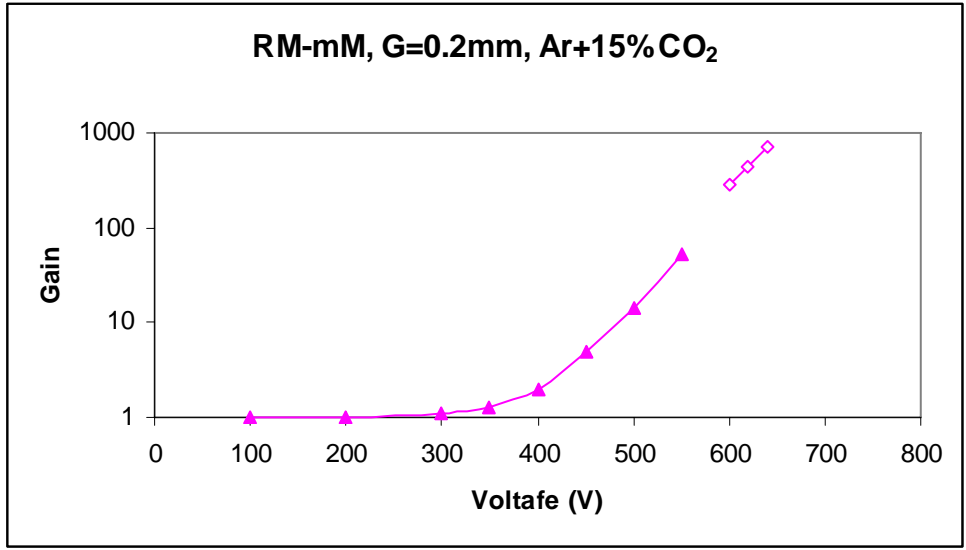

Fig.10. The gain vs. the voltage for RM-µM, G=0.2mm. Symbols represent the same measurements as depicted in Figs. 7 and 8

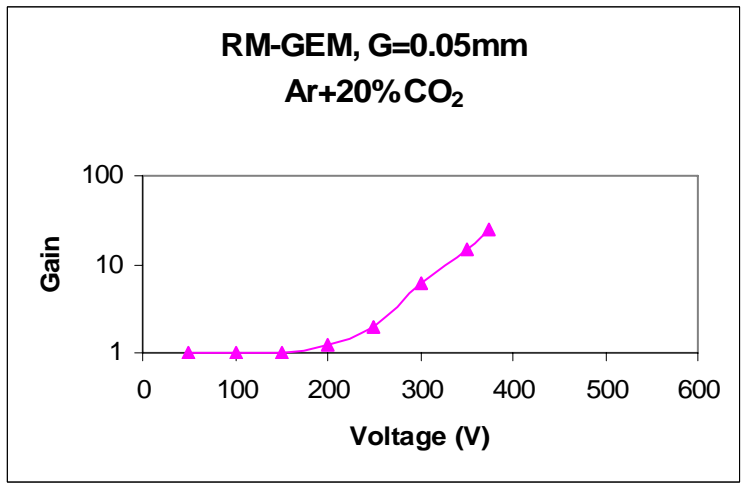

Fig.11. Gain vs. voltage measured for the RM-GEM (Kapton spacers) in Ar+ 20%$CO_2$

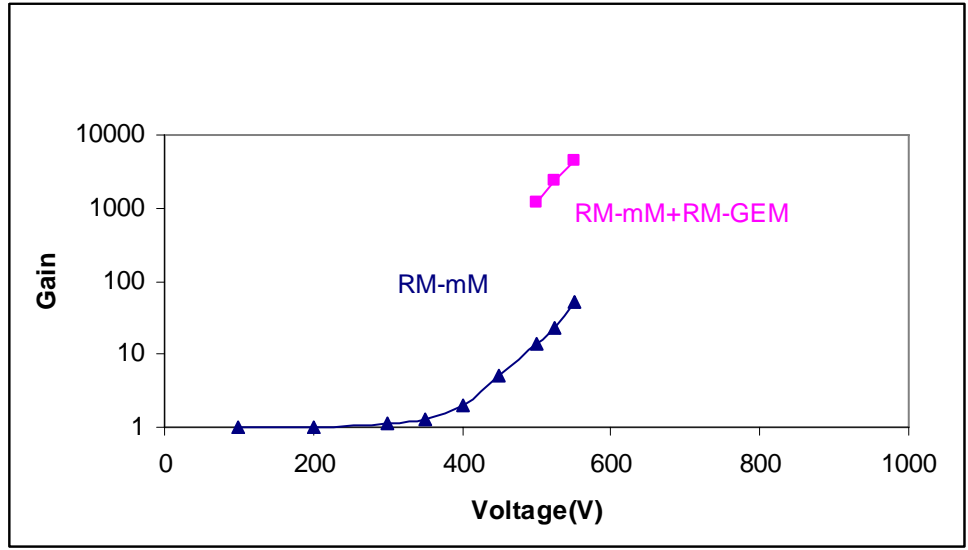

Fig. 12. Results of gain measurements performed with RM-μM ,G=0.2mm (alpha particles) and with cascaded RM-μM combined with RM-GEM,G=0.2mm ($^{55}$Fe source). Gas mixture Ar+15%$CO_2$, voltage drop over RM-GEM 700V, transfer field 1.5kV/cm

3.2 CsI Coated Meshes

As was shown in [4], RETGEM (made of resistive Kapton) with the cathode coated with a CsI layer becomes a photosensitive detector able to detect with a high quantum efficiency single UV photons. Unfortunately, as was already mentioned in the introduction, the position resolution of the RETGEM is quiet modest (0.7-1mm) [13] due to the large hole diameters and their large pitch.
A CsI coated resistive mesh (which has much lower hole pitch than the RETGEM) allows to build small gap detectors which have potentials for much better position resolutions.
To demonstrate this, we used an CsI coated RM- μM with the ceramic anode plate having metallic readout strips. The light from the pulses $D_2$ lamp was focused on the top of this mesh (the diameter of the light spot was about 150 μm). The mean value of the pulse amplitude measures vs. the strips group number is shown in Fig. 13. As can be seen, the width of this distribution at its "half maximum" is about 300μm which is already 2-3 times better that was obtained with a RETGEM. We are quite confident that a much better position resolution can be achieved with a finer mesh and with more accurate measurements and work in this direction is now in progress.

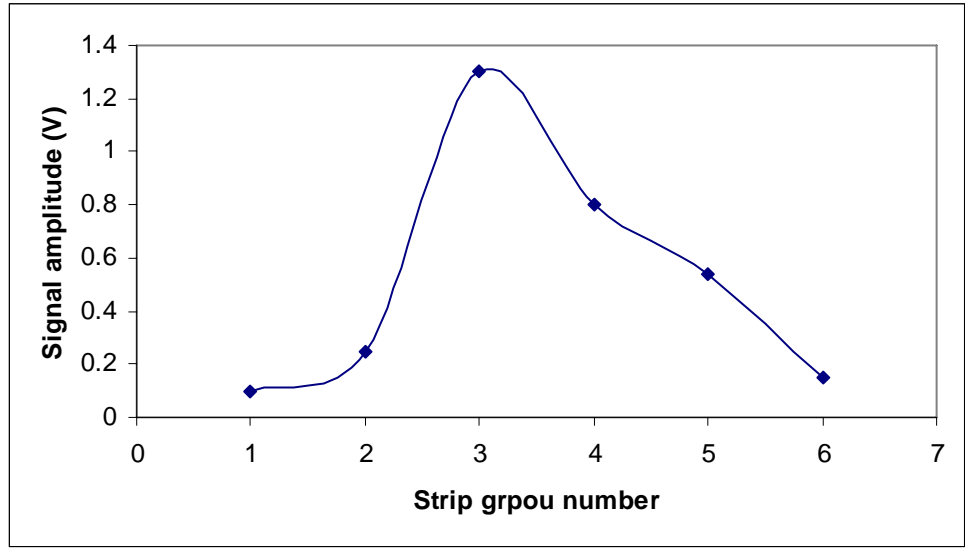

Fig. 13. Signal amplitude vs. the strip group number measures with RM-µM, G=0.2mm in Ar+15%$CO_2$ at a gas gain of 300.

It was also very important to estimate what fraction of the photoelectrons K extracted from the top mesh surface can be drifted to the inner volume of the detector: between the mesh and the anode. This can be rather simply evaluated in the current mode when the detector operates without any multiplication. For this, we measured a photocurrent collected on the drift mesh as a function of the voltage drop between the drift mesh the resistive mesh (in these measurements the resistive mesh was electrically interconnected to the anode plate). Typically, this current increases and the reaches the plateau value $I_d$. We then measured the photocurrent at the anode plate $I_a$ as a function of the voltage between the anode and the resistive mesh at the condition when the voltage drop between the drift and the resistive meshes was kept at zero. For these conditions the collection efficiency can be defined as

$$K=I_d/I_a. \quad (3).$$

Typical results are depicted in Fig.14 showing that even at small voltage at least 80% of photoelectrons are drifted to the anode–cathode gap. It is very probably that at high gains this value will reach 100% (see [14]).

Finally, we estimated the quantum efficiency of the CsI layer evaporated on the top of the resistive mesh. For this the top mesh was irradiated by a185nm light from the monochromator and we measured the photocurrent $I_{a1}$ at this wavelength between the resistive mesh and the drift mesh (Fig. 15). Then the same beam from the monochromator was directed to a single- wire counter flushed with the mixture of CH$_4$ and TMAE vapors and the photocurrent created there was measured at its saturated/plateau value I$_{ref}$. From comparison these photocurrents the quantum efficiency Q$_{185}$ was estimated:

$$Q_{185}=30\% \; I_{a1}/I_{ref} \quad (4).$$

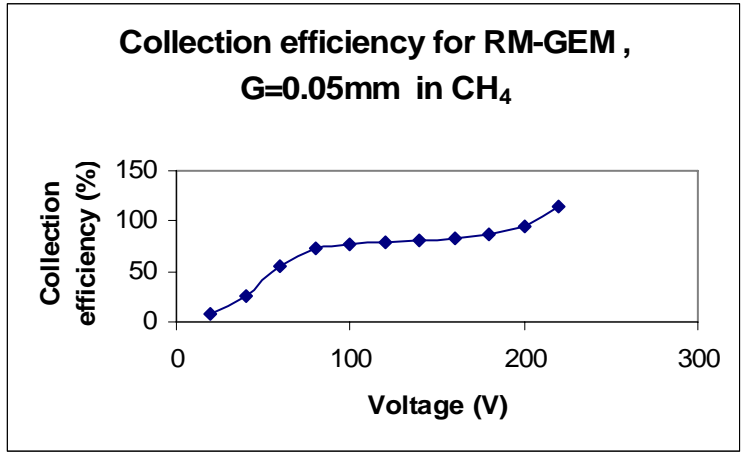

Fig.14. Photoelectrons collection efficiency for RM-GEM, G=0.05mm in CH$_4$. Note that at V>200V the amplification starts, however it is clear that in the voltage interval 100-200V the extraction efficiency is about 80%.

As can be seen from Fig. 15 the ratio I$_{a1}$/I$_{ref}$ is about 0.4 which gives the quantum efficiency value Q$_{185}$~12% at 185 nm.

Thus the resistive mesh can serve as an efficient photocathode for various photodetectors.

Of course, all these conclusions are very preliminary and more measurements are now taking place.

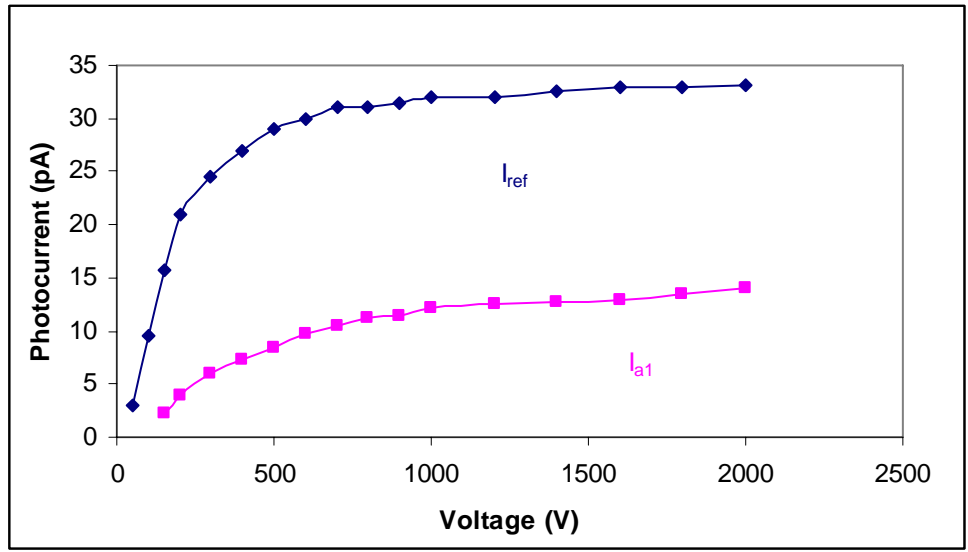

Fig.15. Photocurrents $I_{ref}$ and $I_{a1}$ vs. the applied voltages

## 4. Conclusions.

Resistive meshes developed and tested in this work are convenient construction blocks for various spark-protective detectors including the GEM-type and MICROMEGAS-type.
Due to the small diameter of their holes and the fine pitch, a better position resolution can be achieved with RMDsthan with the RETGEMs.
No discharge propagation was observed in our experiment when RMDs operated in cascade mode. One of the advantages of the cascade mode is the possibility to reduce an ion back flow to the cathode which can be an attractive features for some applications such as photodetectors or TPC.

Our nearest efforts will be focused on developments and tests of fine pitch meshes manufactured by various techniques. This will allow for the building of high position resolution spark protected micropattern detectors. One of the possibilities is to use the fine resistive mesh for MICROMGAS combined with a micropixel readout plate; this approach can be an alternative to current efforts from various groups [8] to develop micropixel anode plate with resistive spark protective coating